# Chinese Interpreting Studies: Genesis of a Discipline

Xu Ziyun

Universitat Rovira i Virgili

# Chinese Interpreting Studies: Genesis of a Discipline[1]

Xu, Ziyun

Universitat Rovira i Virgili

**Abstract/Résumé**

Au cours des deux dernières décennies, on note une forte croissance de la recherche sur l'interprétation en langue chinoise (RIC), comme en témoigne le nombre d'articles publiés. Le présent article adopte une approche scientométrique pour analyser la production, les thèmes et les influences théoriques dans ces articles au fil du temps. Les auteurs, les universités et les régions les plus productifs, de même que les formes de collaboration, ont été analysés afin de permettre une compréhension plus approfondie du paysage de la RIC. Il apparaît que la 'culture' globale de la discipline est restée immuable : aucun de ses influences théoriques et thèmes n'est devenu plus populaire au fil du temps. Cependant, en matière de collaboration entre chercheurs et de politique universitaire, des limites persistent et font obstacle à la réalisation de son potentiel de croissance.

**Keywords/Mots-Clés**

Scientometrics, Chinese Interpreting Studies, research collaboration, theoretical influences, themes

## I. Introduction

The evolution of Chinese Interpreting Studies (CIS) can be traced through the growing number of journal articles and conference proceedings on the subject[2]. Scholars from all over China, from both within and without the universities' established interpreter training programs, contribute to the advancement of CIS research by publishing articles and proceedings. The first CIS article was published in 1958, when Tang Sheng and Zhou Yuliang gave their perspectives on the nature of interpreting, the ins and outs of interpreting competence, and interpreter training. Only two articles appeared in the 1970s, no doubt because of the Cultural Revolution: in that decade nothing was published that was not propaganda, and China cut off its ties with the rest of the world to such an extent that there was practically no need for interpreting of any kind. With its nascent policies of reform and liberalization, 55 articles were published in the 1980s. Since then CIS research has been gaining momentum year on year – 475 articles were published in 2012 alone. This paper seeks to assess the development of CIS by means of a scientometric[3] analysis of one of its most productive sources of research – journal articles and conference proceedings. Using available data from CIS papers, I examine the growth, themes and theoretical influences of those papers over time, as well as the most productive players and their research collaborations – all of which are essential for a comprehensive understanding of the CIS landscape.

---

[1] I wish to thank Daniel Gile for his research guidance. I am grateful to Leonid Pekelis at Stanford University for sharing with me his statistical knowledge of null model analysis and dynamic visualization. I am also indebted to Ewan Parkinson for providing valuable comments on various drafts of my paper.

[2] For the purpose of this research project, journal articles and conference proceedings (many of which are themselves published in journals) are considered as equivalent. A piece of writing appearing in both formats, or in one format in multiple publications, was only collected once for the data-set. Interviews, book reviews, obituaries, reminiscences, discussions of exam questions, and tips for students were all excluded because they represent a different type of data, and are therefore outside the scope of this project.

## II. Previous Scientometric Work

Gile spearheaded scientometric research in Interpreting Studies in the early 1990s, when he started using scientometric data in his research. Early on he began exploring the qualitative elements of citation analysis, indicating the category of each citation (research methods, concepts, terminology, anecdote, etc.) (Gile, 2006). Feeling that citation analysis could offer numerous benefits to students with limited research experience, he taught classes in scientometrics and supervised several masters' and PhD students who applied the method in their research projects (Rowbotham, 2000; Erwin, 2001 & Nasr, 2010). In 1995 he identified trends in Conference Interpreting Studies through scientometric analysis of the relevant literature. At his suggestion Pöchhacker wrote an article for *Target* (1995) in which he analyzed the productivity of individual authors based on the number and type of their published works.

A handful of researchers have applied the principles and methods of scientometrics to Chinese Interpreting/Translation Studies. A few have attempted to provide a broad overview of trends and developments in the discipline by classifying relevant journal articles by theme and giving a few examples of the leading articles in each category (Hu & Sheng, 2000; Liu & Wang, 2007; Li, 2007). Others have gone further by backing up their claims with counts of articles published on a given theme (Mu & Wang, 2009; Tang, 2010).

---

[3] Scientometrics measures the evolution, impact and research production of a given discipline using data from the themes, theoretical influences and research methods found in articles, theses and doctoral dissertations. The approach is useful in Interpreting Studies because it draws together quantifiable data to provide an objective and replicable assessment of a somewhat fragmented field.

Gao (2008) and Zhang (2011) published studies of the similarities and differences between translation and interpreting research in China and the West. While Zhang took a broad view, covering the entire subject of interpreting, Gao's interest was in the cognition aspect of Simultaneous Interpreting Studies. The latter's corpus included articles from eight leading Western academic journals and three premier Chinese journals. Her analysis suggested that recent works in the West were less characterized by debate than was common in the past, and that they frequently drew on findings from contemporary psychology to re-evaluate prior research in the field. Gao noted the strong influence of major Western theories in CIS from the late 20th-century, theories such as Gile's Effort Models, cognitive pragmatics and the Interpretive Theory. She found that, compared to their Western counterparts, Chinese researchers were less inclined to interdisciplinarity.

The effectiveness of scientometric studies depends strongly on their measurements and data quality. In the past Chinese scholars tended either to assess the CIS landscape descriptively and qualitatively, drawing on their intuition and personal experience, or to present conclusions that were based on very limited data and consequently not representative of Interpreting Studies as a whole. In her 2008 work, for example, Gao Bin had a database of 197 papers published in three Chinese academic journals between 1994 and 2007. From this she built a secondary database on SI, selecting a total of 158 citations from just 17 of those papers. Her decision to concentrate on these alone rather than use citations taken at random from all papers on the subject limited her ability to assess larger trends in the wider population. Zhang Wei's paper (2011) serves as a theoretical analysis of the differences in interpreting research between China and the West, but empirical studies need to be conducted to provide data to support or refute his conclusions that there is little innovation in CIS and that most authors concentrate on practical issues such as interpreter training.

## III. The Present Study

In this paper, which is built on the foundations of earlier researchers' exploration of the broad topics studied in CIS, my aim is to take a data-driven approach to further examine the underlying themes, theoretical influences, growth and evolution of CIS, as well as its most productive scholars and regions. I seek to provide answers to the following questions: How has the growth of CIS papers evolved over the past two decades? What are their major theoretical influences and themes, and how do these evolve over time? What research methods are employed? How do scholars study the modes and modalities of interpreting? Who are the most prolific article authors in China? How and to what extent do authors collaborate with one another? Are there any patterns for inter-institutional collaboration? Which regions of China have emerged as the major production centers for CIS research, and what are the patterns for inter-locational research collaboration?

## IV. Methodology

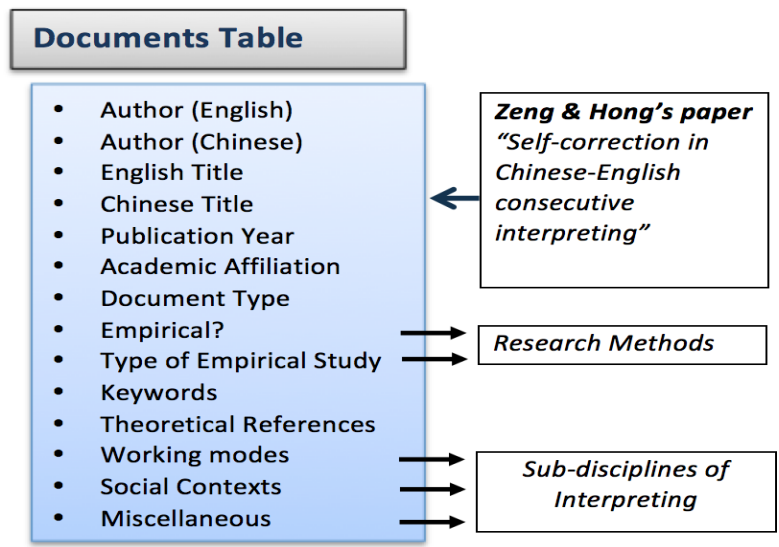

Figure 1: Example of paper labeling method used

The database analyzed in this paper comprises 2,909 journal articles and conference proceedings. To ensure that it was as comprehensive as possible, the majority of journal articles were collected from three sources: CNKI, Wangfang Data and Airiti Library. The conference proceedings were sourced from China's biennial international fora on interpreting from 2006 through 2012. The result was a collection of research papers representative of the whole CIS community. Each paper was labelled[4] for its author(s), title, academic affiliations, research methods[5], keywords, theoretical influences and modes and modalities of interpreting[6].

Once all the data had been collected, it was found that a total of 256 unique theoretical influences and 453 keywords had been logged. To facilitate the isolation of meaningful trends from such a tremendous amount of information, the theoretical influences were grouped into the following six categories:

1) Cognition (Cognitive Science, Psychology, Neuroscience, etc.)

2) Language (Linguistics, Second Language Acquisition, etc.)

3) Communication Theory

4) Translation[7]

5) Peoples and Cultures

6) Miscellaneous (Philosophy, Education, etc.)

In the example of Zeng and Hong's 2012 paper shown in Figure 1, Levelt's Language Acquisition Model belonged to the Language category and Second Language Acquisition sub-category. A coding scheme originally developed by Gile (2000) was adapted with the aim of consolidating the keywords; by this means they could be divided into the following memes[8]: Training, Professional, Language, Socio-cultural, Cognitive and Miscellaneous issues. In the case of Zeng and Hong's paper, both 'self-correction' and 'interpreting performance' belonged to Training issues.

---

[4] For a more detailed description of the labelling scheme, please refer to Xu (submitted).

[5] The papers were divided into the broad categories of theoretical and empirical research. The latter contained the following sub-categories: observational, experimental, interview-based, questionnaire-based, and ethnological studies.

[6] Adapting Pöchhacker's classification system (2004), the different types of interpreting were divided into the following sub-groups: working modes (consecutive, simultaneous, sight translation), social contexts (business, technical, diplomatic, court, religious, signed language, healthcare, conference, escort, community, other) and miscellaneous (telephone and TV interpreting).

[7] Translation theories are those general principles that describe how the practices of both translation and interpreting should be conducted. Given that Translation Studies has become an independent academic discipline with a unique set of paradigms distinct from other fields (Snell-Hornby, 1995), its theoretical influences deserve a separate category of their own.

## V.     Results and discussion

### 5.1 Growth of research papers

Publication count is an important indicator of scholarly activities in a given field (Grbić & Pöllabauer, 2008, p. 91). It can also be used with large data-sets to analyze the research output of a particular region over a long period (for an example see Blickenstaff & Moravcsik, 1982). The number of articles published annually in CIS has grown steadily since the early 1990s; however, the rate of growth has decreased slightly since around 2006. To find a medium-term trend line for the years 1990–2012, a smoothing spline regression was used, with automatic detection of the degree of smoothing by generalized cross validation ($R^2 = 0.99$). This method was chosen because it requires minimal modeling assumptions while fitting a general trend line through noisy data. In other words, smoothing splines are non-parametric (no assumptions are made about the data's distribution), so it gives a descriptive line through the data without confining possibilities to a certain set of models, such as linear regression.

---

[8] The term 'memes' was adopted from Chesterman (1997) to refer to ideas and concepts being replicated and passed down the generations in the manner of human genes.

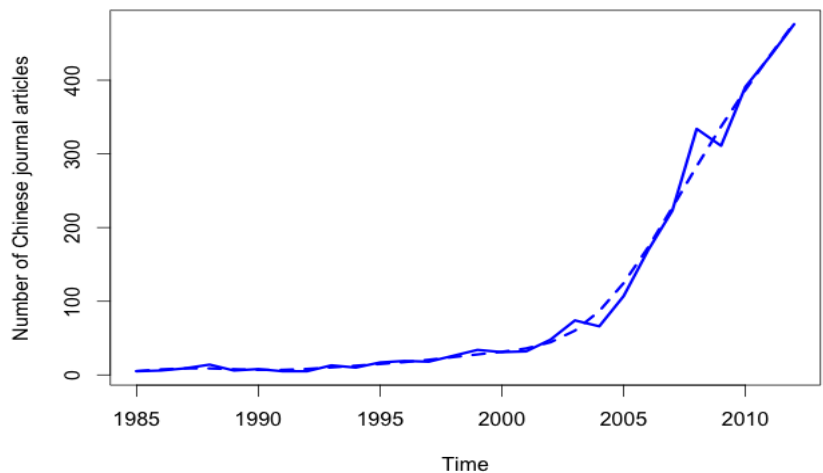

**Figure 2: Research papers published each year in CIS: actual number (solid line) and trend (broken line)**

The sustained growth of articles may be due to the following: there are, at the date of writing this report (October, 2014), 106 universities in China which offer a Bachelor's Degree in Translation and Interpreting (BTI); of that number, 216 have launched Master's Degree (MTI) courses too. In addition, numerous community colleges have introduced courses on business interpreting. Despite the fact that, unlike their Western colleagues, Chinese teaching staff are not obliged to publish in order to gain tenure, their employers nonetheless require them to do so once they have assumed their duties. No doubt their sheer numbers have contributed to the growing momentum of journal publications in China.

### 5.2 Theoretical influences in articles

While publication counts provide a panoramic view of the rapid growth of CIS over the past few decades, they do not reveal its intellectual lineage, i.e. how the ideas and research of preceding scholars have shaped its development. It was with this omission in mind that I set out to study the theoretical influences contained in all the research papers. Theories relating to Cognition were the most popular with a share of 30.4%, followed by Language (21.1%) and Translation-related fields (19.1%). One might have expected Translation Studies to be more dominant given the close relationship between

Chinese Interpreting Studies: Genesis of a Discipline    XU, Ziyun 165

translation and interpreting, but the broad CIS community seems more inspired by Cognition and Language-related fields. Meanwhile, Communication Theory came in last place (6.6%), behind the Miscellaneous (10.5%) and Peoples and Cultures (12.3%) categories.

The most popular sub-category within Miscellaneous was Professional (63%), followed by Education (26%). In the Professional category, analysis of the data revealed that the majority of authors drew on theories relating to various interpreter training models, such as Xiada's (64 articles) (Lin et al., 1999), Guangwai's (61) (Wang, 2009) and AIIC (7)[9]. In the Education category the data indicate that the majority of articles (40) used pedagogy as their theoretical foundations; the most popular education theory, that of Learning Motivation (Stipek, 1988), received only six mentions. In addition, it was found that the third most popular sub-category of Miscellaneous was Philosophy (11%). Eleven articles talked of that subject in general terms, but nine used Wittgenstein's theoretical framework (Wittgenstein, 1953) to argue that interpreting can be thought of as a game and all its participants as players.

### 5.2.1 Most popular theories in CIS

For a more nuanced analysis a careful examination was made of the individual theories associated with each paper in order to identify their influence in the literature over time. Those most frequently mentioned were: the Interpretive Theory of Translation (281 mentions), Effort Models (224), cross-cultural communication (190), cognitive psychology (114) and the Comprehension Equation (74). Because some authors did not refer to specific theories, I was obliged to resort to the use of some rather generalized categories in this list, such as cross-cultural communication and cognitive psychology. Nonetheless, certain individual theories did stand out from the crowd:
- The Interpretive Theory of Translation. This theory has enjoyed great popularity in the history of Interpreting Studies since being jointly developed by Danica

---

[9] The Xiada, Guangwai and AIIC models mainly address how interpreters should be trained, as opposed to offering theoretical explanations of their thought processes or providing specific strategies for dealing with translation issues. For this reason they were grouped under Professional, whereas Gile's Effort Models, which offer suggestions on how interpreters can best use the limited mental resources available to them, were tagged as Translation and Cognitive Science.

Seleskovitch and Marianne Lederer (Seleskovitch, 1978). Despite the flaws in some of its assumptions (Gile, 2014, Bulletin 47), it has served as the theoretical guiding principle for a significant number of Chinese research papers.

- Effort Models & the Comprehension Equation, as developed by Daniel Gile (1995). The former addresses the allocation of mental resources, while the latter illustrates the essential skills and knowledge required for comprehension in interpreting. A great number of Chinese authors have used the two theories to explain the challenges faced by interpreters or to develop coping tactics for improving interpreting performance.
- Cross-cultural communication and cognitive psychology. The fact that these two broad disciplines inspired a large number of CIS scholars in their research indicates that they fully appreciate the advantages of interdisciplinary work, but their habit of making non-specific references (citing works rather than particular theories) suggests that they are not completely at ease in these 'foreign' fields.

To present a full picture of the theoretical influences on CIS, a list of moderately popular theories was selected: Schema Theory (73 mentions), Relevance Theory (64) and pragmatics (52) (Leech, 1983).

- Schema Theory. With its origins in psychology and cognitive science, a number of scholars used this to address topic familiarity, background knowledge, working memory and interpreting quality.
- Relevance Theory. First proposed by Dan Sperber and Deirdre Wilson (1986), the popularity of this theory highlights scholars' belief in the importance of implicit information in communication. Chinese is rooted in a high-context culture[10] (Hall, 1976), in which many things left unsaid can still be understood by people of the same background. Translating Chinese into English, which is rooted in a low-context culture, can be problematic for interpreters, who are tasked with unpacking the hidden meanings between words.

---

[10] A high-context culture is one that uses the common and shared experience of the group to communicate complex messages in few words, whereas a low-context culture relies more on the individual words themselves to get the speaker's ideas across.

- Pragmatics. This branch of linguistics is mainly concerned with how context influences meaning. Analyzing instances of pragmatic failure is a common approach taken by Chinese scholars to address errors and issues that arise from cross-cultural communication.

### 5.2.2 Consistency of theoretical influences on CIS

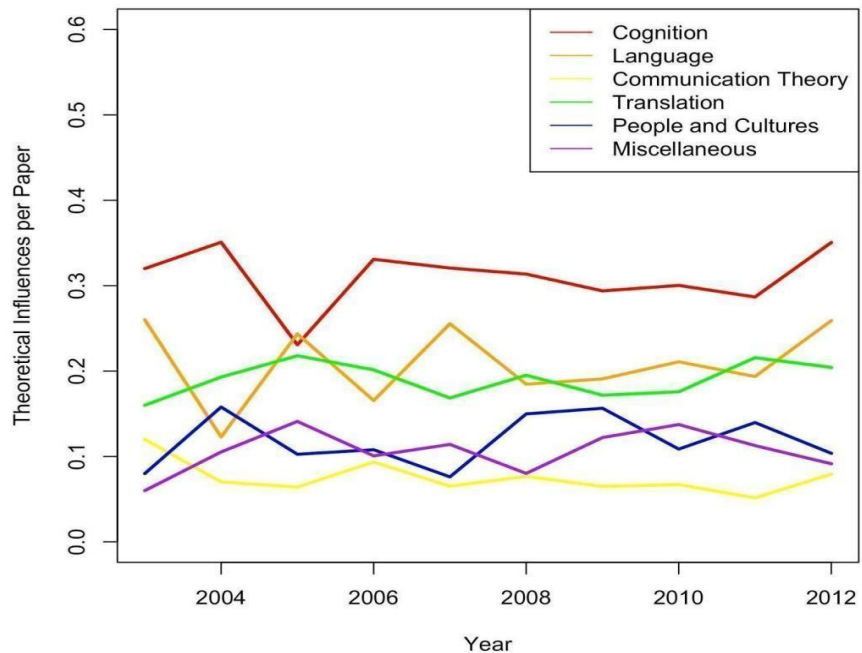

**Figure 3: Average number of theoretical influences of each type per paper over time**

A time-series analysis of theoretical influences on CIS from 2003 to 2012 was conducted to determine the presence or absence of trends (see Figure 3). This time frame was selected because very few articles (2.86% of the total) were published before 2003: the theoretical influences recorded in them accounted for only 5.98% of the overall total. The fluctuating trend lines for various influences are immediately noticeable: 1,161 of the 2,909 articles (almost 40%) in the data-set lacked identifiable theoretical influences, which led to the low proportions for each category; this in turn led to the trend lines being more prone to fluctuation. There are two possible explanations for this paucity of theoretical influences in CIS papers: 1) article authors did not have the necessary space to flesh out the theories that inspired their work; and 2) the trend for drawing on existing theories did not really begin until 2003: before then papers were more likely to take the form of descriptive studies of data-sets. Despite the fluctuations in the data, Figure 4 would seem to suggest that influences from different theories were generally stable, with Cognition-related disciplines being the most prominent, while Communication Theory and the Peoples and Cultures categories remained rather flat and low.

To determine whether the proportions of different theoretical influences did indeed remain constant over the displayed time period, a multinomial logistic regression model[11] was fit, using the publication year as a predictor.

**Table 1: Analysis of Deviance table for multinomial logistic regression of theoretical influence proportions**

|      | LR Chisq | Df | Pr(>Chisq) |
|------|----------|----|-----------|
| Year | 1.81     | 6  | 0.9367    |

To test whether theoretical influences vary from year to year, a horizontal line was fit to the average number of times an influence was referred to in a paper. It was then possible to see if there was a difference between those lines and the ones seen in Figure 4. After this fit was completed, an Analysis of Deviance was performed: its p-value was found to be 0.937 , indicating that publication year has no discernible effect on the proportions of theoretical influences. In other words, it is reasonable to conclude that theoretical influences have not changed substantially over the years.

A possible explanation for the consistently low proportions of the Communication Theory and Peoples/Cultures categories is that all interpreting courses in China are offered by schools/departments of translation or foreign languages, as opposed to faculties of intercultural studies or communication – indeed, the Ministry of Education lists Translation Studies as a branch of Foreign Languages and Literature in its classification of disciplines. This institutional ethos inevitably limits students' exposure to cross-cultural and communication studies.

---

[11] Multinomial logistic regression is a generalization of the more well-known binomial logistic regression. The latter involves the modeling of binomial data in which a certain number of successes in the total number of trials are observable. The former, by contrast, models the probability that one of the observations is in each of the $k$ groups. Thus the sum of the proportions of successes in each of the $k$ groups will always add up to one. As each of the influences observed in a given work must of necessity fall into one of the six categories, the data for the theoretical influences of CIS research papers can be modeled by means of multinomial logistic regression.

## 5.3   Memes of CIS

If theoretical influences can be considered the 'input' which inspires CIS researchers to conduct their studies, memes are the 'output'. The collected data suggest that Training was by far the most popular meme in research articles, with 51.2% of the total. Memes such as Cognitive (15.8%), Language (12%) and Socio-cultural issues (10.7%) enjoyed modest levels of popularity.  Only a small proportion of authors addressed Miscellaneous (5.1%) and Professional (4.9%) issues. It should be stressed that the majority of the article authors are academics who hold teaching positions at universities; most have not served as advisors to graduate students (see Section 7). The fact that a large number of authors write about Training could be due to a number of factors such as personal interest, familiarity with the topic, or convenience; however, the reason may simply be that there exists in China an urgent and overriding need to find the most effective ways of transforming students on all the newly created BTI and MTI programs into qualified interpreting professionals. More research needs to be done to pinpoint precisely what motivates scholars to write more about Training than any other subject.

It was rather surprising to see Professional issues receiving scant attention in journal articles. This relative unpopularity can perhaps be ascribed to the fact that most Chinese article authors do not work actively as interpreters; by contrast, these issues received a great deal of attention in Western publications, the majority of their authors being working professionals (Gile, 2000).

## 5.3.1 Keywords in CIS

A fine-tuned analysis was conducted to reveal the most frequently mentioned keywords. The top three were rather general: interpreter training (987 mentions), interpreting strategies (559), and interpreting techniques (252). However, a few others deserve our special attention — these highlight more specific research areas in Interpreting Studies:

- Note-taking (242 mentions): an important component in consecutive interpreting, authors have analyzed this particular skill from various angles, e.g. Effort Models, Relevance Theory (Carston & Uchida, 1998), and the choice of language in which notes should be taken.
- Curriculum design (201): following the establishment of the European Masters in Conference Interpreting (EMCI), scholars have realized the need for a uniform teaching plan to make up for the shortage of qualified instructors in certain areas of China.
- Interpreting textbooks (167): Lv (2001) points out that a large percentage of Chinese interpreting manuals are out of sync with the latest developments in educational theory — for this reason the creation of pedagogically sound books has become a popular area of research.

A few moderately popular keywords offer a different perspective on the CIS landscape:
- Online learning (99): given that large class sizes can cause problems for MTI and BTI instructors, not to mention those teaching at community colleges, a lot of Chinese universities have been pushing hard for the introduction of self-managed online learning. In addition, simple love of the latest and most on-trend technologies may be another driving force behind research into online learning.
- Background knowledge (87): some authors used Schema Theory (Arbib, 1992) to address the impact of background knowledge on overall interpreting quality.
- Context (77): other authors chose to investigate the role context plays in listening comprehension and anticipation, and its effect on interpreting performance.

### 5.3.2 Consistency of memes in CIS

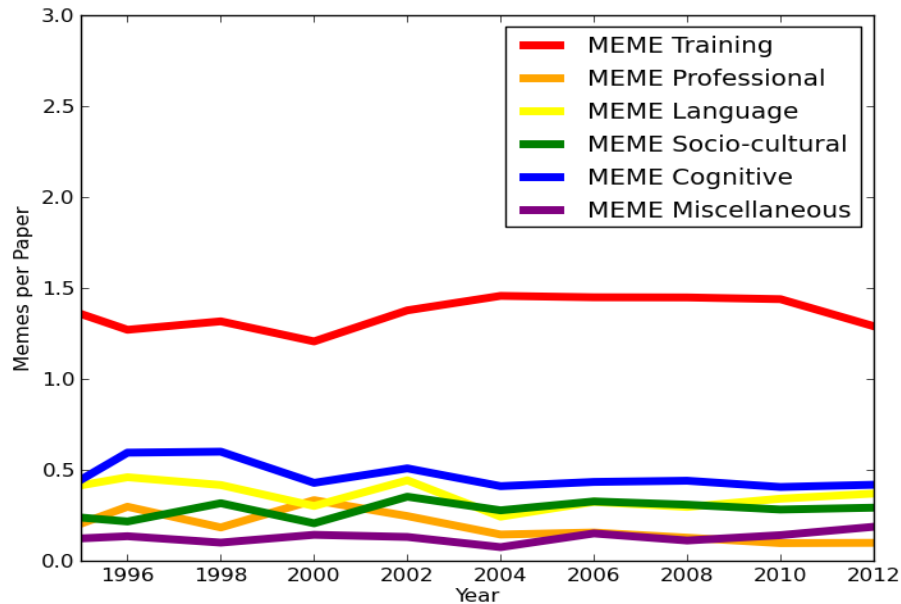

Figure 4: Average number of memes of each type per paper over time

None of the memes in research articles grew appreciably in popularity over time (see Figure 4). Training was by far the most frequently discussed meme (of 2,909 articles, 2,378 addressed training), followed by Cognitive (935 articles), Language (826), Socio-cultural (737), Miscellaneous (375) and Professional issues (342). This consistency indicates that CIS has stabilized over time, the most popular themes having remained broadly the same over the last ten years. On examining Training more closely, it was observed that 194 articles dealt with undergraduate-level instruction and 82 with business interpreter training at community college level. These numbers clearly reflect the fact that instruction in interpreting, which is the domain of post-graduate education in the West, is considerably more popular at all levels in China.

### 5.4 Empirical research in CIS

While a number of authors have studied the growth of CIS and investigated the wide range of topics it covers (e.g. Hu & Sheng, 2000; Gao, 2008; Tang, 2010), few have looked into the research methodologies employed by its academics. Examining the methodological trends in a given discipline can help improve the ways in which its research is conducted (Liu, 2011). Of the 2,909 articles in the data-set, 533 (18%) used empirical methodologies; of these, 286 used observation, 151 questionnaires, 148 experiments, and 53 interviews. None dealt with questions of ethnology. A possible reason for the low proportion of empirical studies is that few authors can afford to spend a great deal of time collecting data for the sole purpose of publishing a single article. Also worth noting is that questionnaires were more commonly used by professors than by students, often with a completion rate of more than 95% – no doubt teachers find it easier than students to persuade large numbers of people to complete surveys!

### 5.4.1 Variation over time of empirical research

Prior to 1994, with the exception of nine articles, all the Chinese authors explored interpreting techniques and competence from the perspective of their personal experience. However, things have moved on from those days: the proportion of empirical research has been growing steadily despite the fluctuations in the late 1990s, which were likely caused by the limited number of articles published at that time. It is obvious from Figure 5[12] below that the sharply rising proportion of empirical papers – up to 30% in 2012 from virtually zero in 1994 – has gone hand in hand with a progressive rise in the use of all four research methods. In particular, the growth of observational and experimental studies has been impressive. The first type saw consistent growth after 1998, and increased to 15% of the empirical total in 2012. Experimental papers, on the other hand, only began to be written in 2000, and have since grown to be more than 5% of the total. With the expansion of MTI and BTI education in China, it has become easier for authors to recruit student participants for their observational and experimental studies, but it would be interesting to see how the results from those studies might be extrapolated to the interpreter population at large.

---

[12] A cumulative plot was created because it clearly shows how both the overall trend (the proportion of empirical articles) and the components of that trend (the proportions of the four different empirical research types) fluctuate at the same time. In other words, it allows us to compare subcategories over time and to see how each contributes to the growing proportion of all empirical papers over the same period.

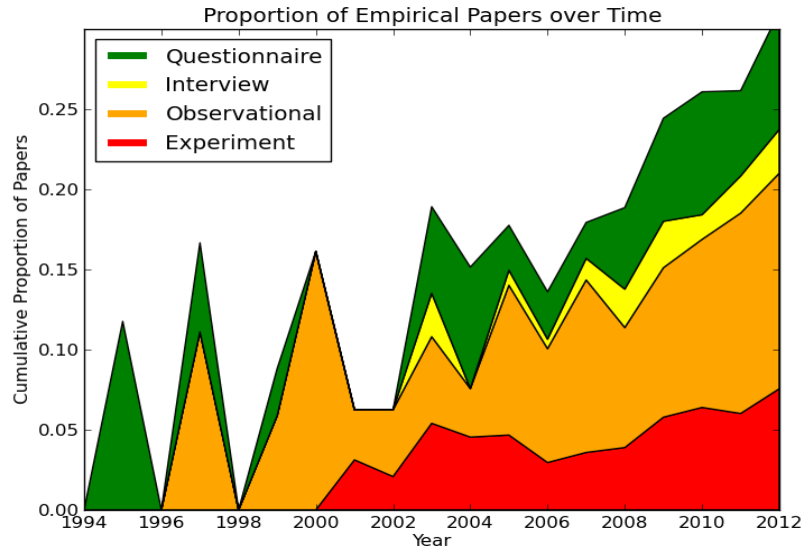

Figure 5: Cumulative proportions for different categories of empirical research over time

## 5.5 Modes and modalities of interpreting in CIS

Interpreting — a social activity whose purpose is to facilitate communication between two parties of different linguistic and cultural backgrounds — is far from being a homogenous entity. Many Western scholars specialize in one of its varieties: Daniel Gile and Robin Setton, for example, concentrate on conference interpreting, while Mary Phelan specializes in community interpreting; Franz Pöchhacker and Miriam Shlesinger cover both these. In China all interpreter training programs focus on producing conference interpreters, so it would be interesting to see if this is reflected in CIS research papers.

When the working modes of interpreting were examined, it was found that 736 papers (25.3%) addressed consecutive interpreting (CI), making it the most frequently discussed mode, followed by 244 (8.4%) for simultaneous interpreting (SI) and 66 (2.3%) for sight translation (ST); the remainder did not specify the mode investigated.

Unlike the interpreting market in the West, where SI is the working mode preferred by employers (Gile, 2000), CI dominates in China's major translation markets (Pan, Sun, & Wang, 2009; Wang, 2006); this undoubtedly goes a long way to explaining its popularity as a research subject. Also, the fact that students are not exposed to SI at all at BTI and community college level no doubt has a bearing on the figures.

When it came to examining social contexts, it was observed that Business Interpreting received the most attention (40), closely followed by Diplomatic (37) and Conference Interpreting (30). Escort (21), Technical (16), Healthcare (13), Court (10), Signed Language (7) and Community in general (2) also received modest amounts of attention. Furthermore, of the more than 2,900 papers fewer than 100 explicitly stated that they were investigating a specific social context. This suggests that the majority of authors do not feel it necessary to address a specific context; rather, they tend to talk about interpreting broadly and holistically. Business interpreting received the most attention from authors of papers: this was perhaps due to their working at community colleges where students are trained to facilitate investment talks and trade negotiations rather than to work in diplomatic or conference settings.

Aside from the aforementioned contexts of interpreting, it was observed that 16 articles addressed TV interpreting and 14 telephone interpreting. These small shares of the total indicate that Chinese scholars have accorded them scant attention, and that they are as a result relatively under-researched in comparison with the other types. TV interpreting is used when foreign guests are invited to participate in live shows, or important speeches are made overseas. Its lack of popularity as a research subject can perhaps be ascribed to established conference interpreters' antipathy toward it: firstly they resent the fact that TV news reporters with no training in SI often take on the task, causing potentially serious damage to the reputations of professionals (Zhou, 2007); and secondly, when called in for TV assignments, they often find the stations sorely lacking in the proper equipment and working conditions (Wen, 2006). Telephone interpreting is primarily used in countries with large numbers of immigrants such as the United States and Australia but has started to gain prominence in China, as was seen, for instance, when the country hosted the 16th Asian Games and the 1st Asia Para Games (Zhan & Suo, 2012). As China becomes increasingly interconnected with the rest of the world via commerce, politics, sport, and other cross-border activities, it will be interesting to see if these two types of interpreting increase their share of the market and gain popularity among Chinese researchers.

### 5.5.1 Variation over time of working modes

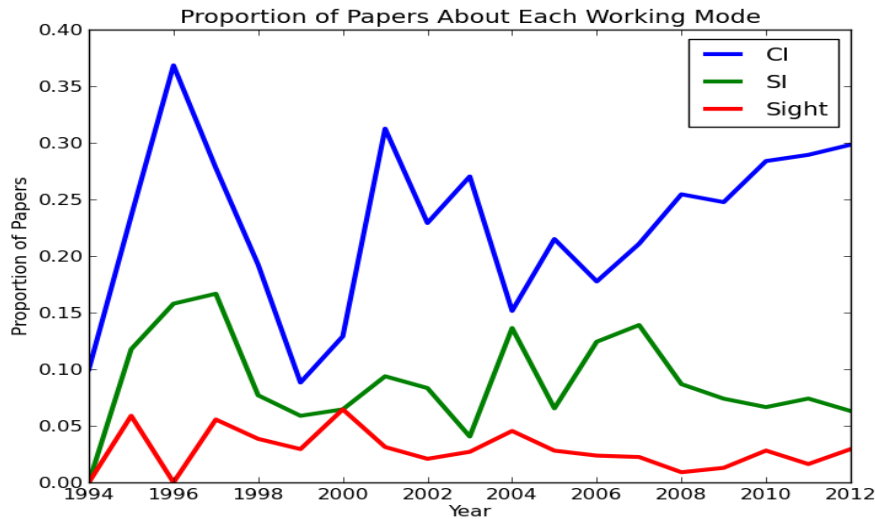

**Figure 6: Proportions of papers dealing with working modes of interpreting over time**

Given that China's interpreting market has changed so dramatically over the past two decades, I thought it would be worthwhile to investigate whether its scholars' research interests had changed with it. With this in mind, the data was plotted over time, as shown in Figure 6. CI was the most studied mode over the period; the popularity levels of the three modes remained relatively steady[13]. It ought not to be forgotten that the start of CI's steady increase in popularity in 2006 coincided with the Ministry of Education's introduction of the BTI: three universities were chosen to pilot the course in that year, with the total number gradually expanding to 106.

It is also worth mentioning that ST attracted consistent – if only moderate – attention. There are a few possible reasons for this situation: article authors, most of them interpreting instructors, may have realized this mode's importance in helping students develop the skills required for SI. Given that both BTI courses and interpreter training at community college level are enjoying booming popularity in China, the subject of ST, along with CI, has been introduced into many more classrooms than SI, which is mainly reserved for post-graduate study.

---

[13] It should be noted that the yearly values in Figure 6 do not add up to 100% since a large proportion of papers did not focus on a specific mode but dealt with interpreting in general.

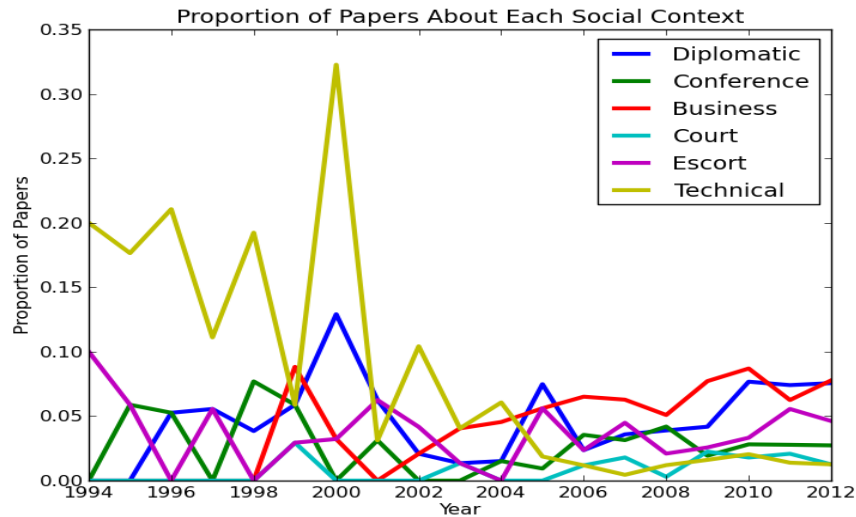

**Figure 7: Proportions of papers dealing with social contexts of interpreting over time**

It became clear in examining the trends for the social contexts of interpreting (see Figure 7) that there is greater fluctuation here: this is no doubt due to the limited number of articles that specifically address these topics, which makes the entire picture rather messy. However, it is immediately noticeable that technical interpreting, which enjoyed considerable popularity in the past, has witnessed a sharp decline since 2001. Changes in China's interpreting market may be at the root of the falling number of research articles on this topic. There was a time when, because the majority of the country's engineers spoke limited English, technical interpreting was a mainstay of the market, an indispensable adjunct to the large number of foreign technology transfers between China and the rest of the world. Over the past decade the number of nationals who have worked or studied abroad has increased sharply, so that many of the functions that were once the preserve of qualified career interpreters have now been taken up by in-house staffers, who find themselves at a triple advantage: they have the technical know-how, they are proficient in English, and their first allegiance is to their companies.

## 5.6 Most productive authors

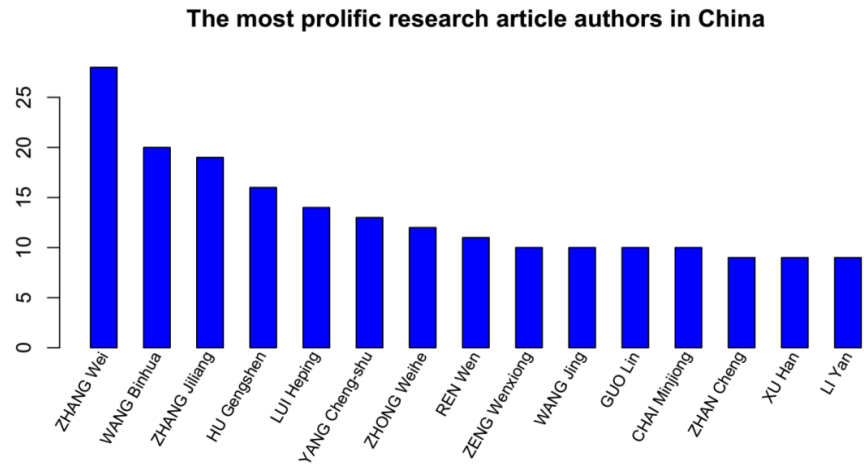

**Figure 8: Most prolific authors of research articles in China**

Of course the dynamic evolution of CIS could not have taken place without the contributions of individual scholars. To assess the most productive in the field, the papers of each of the authors in the data-set from 1958 to 2012 were tallied up, one credit per paper being awarded to each author, regardless of the number of co-authors he might have had — Figure 8 shows the results[14]. It came as something of a surprise to find that the most prolific of all, Zhang Wei (28 published articles), in fact majored in written translation[15]: translation and interpreting require distinct skills sets, and there is little doubt that many more academics specialize in just one field rather than expending equal energy on both. A corollary observation is that of the top ten, only Zhang Jiliang and Liu Heping work actively as interpreters, indicating that CIS researchers are not necessarily working interpreters. It is quite possible that these authors without professional experience are motivated to publish because, as interpreter trainers, they are required by their schools to produce a certain amount of job-related research per year.

---

[14] No time-series analysis was conducted to see how the ranking of top producers changed over time because the most prolific authors all had fewer than 30 publications each. Conducting a time-series analysis would involve so few papers per year that only severely limited conclusions could be drawn from the results.
[15] Zhang's graduate thesis (1998) dealt with strategies for translating long English sentences.

### 5.6.1 Co-authorships

The first co-authored research paper appeared several centuries ago in 1665 (Beaver & Rosen, 1978). Thanks to the surging popularity of the Internet and ongoing development of telecommunications infrastructure, both of which have made remote research collaboration increasingly practicable, the number of co-authored papers in all academic journals worldwide has been rising steadily in recent decades (Persson, Glänzel, & Danell, 2004); this is particularly the case for the social sciences, which rely heavily on collaborative research (Endersby, 1996). Cho et al. (2010), in research based on all the Chinese articles archived in the ISI database, a leading provider of bibliographic information on citations, point out that co-authorships have a positive effect on the number of articles published. Given that the majority of publications in interpreting studies are not archived by ISI, I expanded on Cho's findings by investigating co-authorship – this being the most important indicator of research collaboration (Melin & Persson, 1996) – in all 2,909 of the CIS research papers in the data-set.

Of all those articles and proceedings, 2,345 were written by a single author, 495 by two, 63 by three, 3 by four, 2 by five, and 1 by six[16]. In contrast to the mainstream trend in the field of social sciences, in which multiple researchers routinely bring their distinct skill sets to the table to solve complex problems together and where co-authorship rules the day (Endersby, 1996), the majority of CIS articles are authored by just one person. It is clear from the numbers given immediately above that single-author is the format of choice among CIS authors, leading one to wonder if Chinese researchers might be somewhat insular.

---

[16] This six-author article was written by graduate students and faculty members from the Beijing International Studies University, who investigated the effect of input speed on the completeness of rendition in CI. It is interesting to note that the number of research participants was the same as that of researchers.

## 5.6.1.1 Variation over time of co-authorships

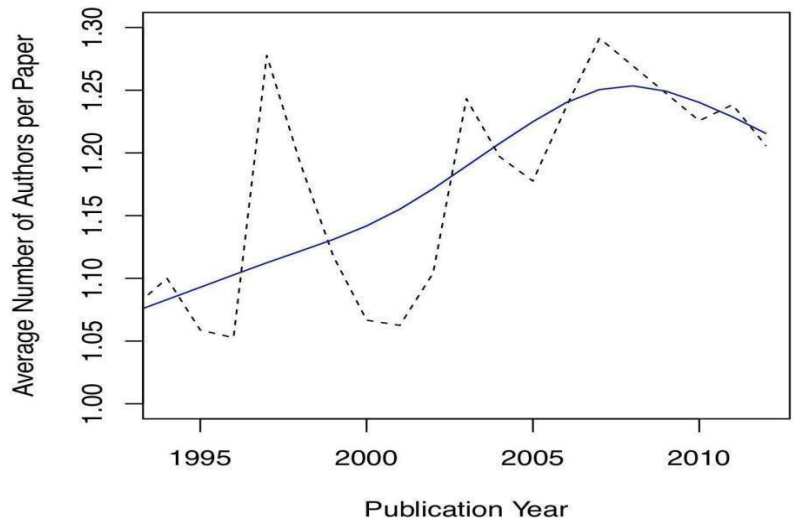

**Figure 9: Average numbers of Chinese co-authorships per year over time**

To understand the trends in co-authorship, I fit a trendline via a smoothing spline; similar to the fit for the growth in the number of Chinese papers, I used R's leave-one-out cross-validation that is built in to the smooth spline function to determine an optimal smoothing parameter value (lambda=0.4059551). Owing to the limited data available for the period before 2000, there was great fluctuation in the average number of co-authors (see Figure 9). Starting from 2000, it increased from 1.06 to somewhere between 1.2 and 1.25 in 2012. However, there was clearly a somewhat downward trend in the average number of authors per paper starting in 2007/2008. In China's liberal arts academic community, an author needs to be the first-listed for a work to count towards meeting his school's research target and for him to be considered for promotion to an associate professorship[17]. Academics may be very reluctant to collaborate with their colleagues because of this administrative constraint.

## 5.7 Inter-institutional and inter-locational collaboration

---

[17] Associate and full professors can be listed as other than first authors to meet their research duties (Shanghai International Studies University, 2007).

Collaborative research is seen as a sign of the increased professionalization of a discipline (Beaver & Rosen, 1978). It offers an opportunity for different institutions to share resources and the technical expertise of their staff who may otherwise only have the time and energies to produce parts of papers. The data-set contained 218 papers with authors from multiple universities (207 had authors from two universities, and 11 from three), from the overall total of 564 co-authored papers. So some 37% of co-authored papers were inter-institutional. This high percentage of collaboration between universities indicates that a significant number of CIS scholars have established their own research networks, to better capitalize on one another's strengths and resources.

With the advent of Internet and other communication technologies, it has become much easier for researchers to collaborate instantaneously, via email and video conferencing systems, unhampered by geographical distance. With this in mind I also decided to examine collaboration between authors from different locations (cities, provinces, regions, countries). Firstly articles with more than one author were collected; their authors' universities were identified and locations noted; and finally all the various pairings were listed. An intra-locational relationship is one in which both (or all) authors come from the same place (e.g. 'three authors from Anhui' = one intra-locational pairing); whereas in an inter-locational pairing the authors come from different places (e.g. 'three authors, one each from Anhui, Beijing and Chongqing' = three inter-locational pairings: Anhui-Beijing, Anhui-Chongqing, and Beijing-Chongqing). The following chord diagram (Figure 10)[18] was created from the resulting calculations.

---

[18] For a more dynamic visual effect, readers can visit the permanent link hosted on my website: http://www.interpretrainer.com/journal_regions_chord.html. Hovering the cursor over the edge of the circle or the key highlights each region's co-authorships. The peak for each region represents intra-locational collaborations, while the linking chords represent inter-locational ones. The width of each peak and chord corresponds to the number of pairings.

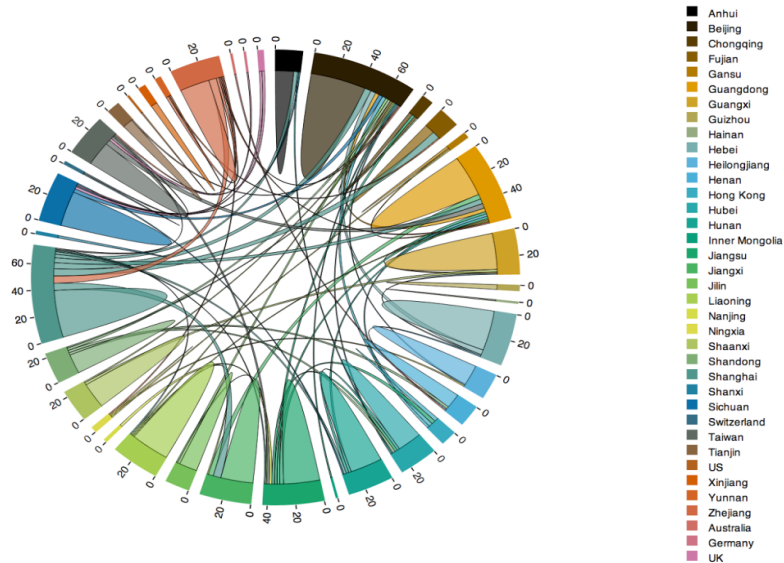

**Figure 10: Inter-locational co-authorships of articles and proceedings**

The exercise revealed 632 co-authorship pairings between locations. Beijing and Shanghai led the way with 74 collaborations apiece, followed by Guangdong with 59. Of Beijing's total, 43 were between authors within the city itself and 31 with authors from other locations. Of Shanghai's total, 39 were intra- and 35 inter-locational; the numbers for Guangdong were 34 and 25. In general, the level of inter-locational collaboration across China was rather low. From this it can be surmised that a significant majority of authors prefer to work face to face with colleagues from similar academic backgrounds (fellow alumni, for example) with whom they have been able to build relationships of confidence and trust over time. The vast majority of collaborations were between Chinese authors, a mere four having been co-authored with overseas colleagues (UK - 2, Australia - 1, Germany - 1).

Gile (2013) points out that international collaboration is particularly beneficial to research because not all countries share the same resources and areas of expertise. In addition, researchers from different countries can conduct experiments using a single piece of source material (one in English, for example), having it interpreted into different languages using the same research methodology. This international approach can help address the problem of language specificity and provide much-needed data in various languages. With China's ongoing biennial conferences on interpreting studies attracting scholars from numerous different countries, an increased level of research across regions and countries seems a likely prospect.

## 5.8  Production centers

As was mentioned previously, interpreter training has become increasingly popular all over China since the introduction of the MTI and BTI degrees in 2007; these in turn have led to an increase in the production of interpreting literature. For this section I examined the data to find out whether research clusters have developed for Chinese journal articles and proceedings. After the authors' academic affiliations had been grouped according to place of origin, they were then plotted on a map of China. The map was generated using a web application based on the 'R' statistical software and its 'shiny', 'maps', 'maptools' and 'sp' packages, all of which are available for free on R's comprehensive archive network, CRAN. Once article counts per location were obtained, this tool allowed for their straightforward graphical representation (Figure 11) – the darker the shade[19], the higher the number of articles. The tools can be found on http://spark.rstudio.com/yichuanw/ChinaMap/.

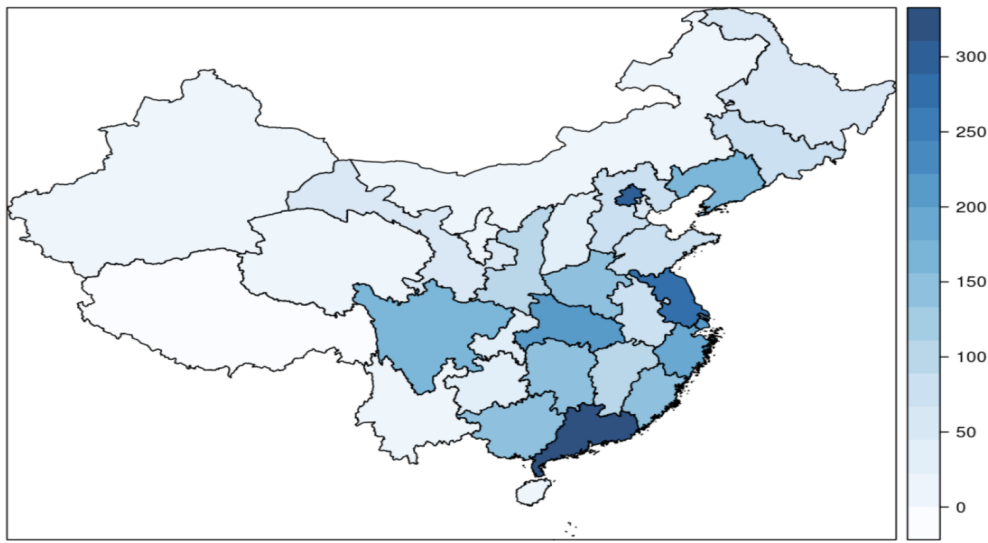

**Figure 11: Research paper distribution across China**

---

[19] For a full-color version of Figure 11, please visit http://interpretrainer.com/image10.png

From Figure 11 it is plain to see that most of the research was concentrated in Southeastern and Central China, the most productive locations being Guangdong (311 mentions), Beijing (309), Jiangsu (265), Hubei (211) and Shanghai (207). The country's coastal regions have enjoyed high levels of economic productivity and developed intensive industrial and commercial activities, which have no doubt led to more international contacts and a higher demand for business interpreters. A case in point is Guangdong province, where the total number of papers published far exceeded that of Beijing which, despite its strategic and political importance as the capital of China and its high concentration of first-tier universities, is landlocked and not a trading hub. In addition, the Yangtze River Delta, another important economic hub for foreign trade, comprises three more major contributors to CIS research: Shanghai (207 mentions), Jiangsu (265) and Zhejiang (198). At the other end of the scale, economically less-developed regions such as Xinjiang (17), Inner Mongolia (14) and Yunnan (17) contributed very little to the CIS literature, despite the existence of multi-lingual cultures and ethnicities in these regions.

Research into interpreting, which is often funded by institutions, is not a lucrative endeavor. Furthermore, Chinese researchers typically have to pay an article processing charge when their works are accepted by academic journals; these range from a few hundred to several thousand dollars depending on the journal's ranking. Both these factors may deter scholars in financially disadvantaged areas from actively carrying out and publishing research, the funding they receive from their schools being insufficient to cover the two expenses.

## VI. Conclusion

In parallel with its economic and trade liberalization, China has rapidly emerged as a large-scale producer of research articles on interpreting studies, and the discipline's growth remains strong even after a decade of massive expansion. Cognition remains the dominant theoretical influence in CIS, followed by language- and translation-related disciplines. Training remains the most studied topic, its popularity perhaps having been driven by the creation of numerous MTI and BTI programs throughout the country. The general 'culture' of CIS remained constant throughout the period, none of its theoretical influences or memes having gained significantly in popularity.

The proportion of papers of an empirical nature is still rather low at only about 20%, in contrast with 50% in MA theses (Xu, 2014)[20]; however, the proportion has been rising over time. CI remains the most frequently studied mode, with its popularity on the increase. Single-author research is the norm in CIS; the limited number of collaborative research articles are produced in discrete areas of the country - few are inter-regional, still fewer international. Furthermore, most research remains concentrated in a small number of the more economically developed areas.

China has blazed a trail by recognizing Translation and Interpreting Studies as an independent discipline (Gile, 2010). It is to be hoped that remote collaboration between universities and regions will become ever more attainable with the aid of modern technologies, giving rise to research which is more representative of the overall interpreter population. No doubt changes in certain academic policies, such as recognizing the research contributions of second or third authors in article publication, would help accelerate this trend.

# References

Arbib, M. A. (1992). Schema theory. In *The Encyclopedia of Artificial Intelligence* (Vol. 2, pp. 1427-1443). Information Science Reference.

Beaver, D. D., & Rosen, R. (1978). Studies in scientific collaboration. *Scientometrics, 1*(1), 65-84. doi: 10.1007/BF02016840

Blickenstaff, J., & Moravcsik, M. J. (1982). Scientific output in the third world. *Scientometrics, 4*(2), 135-169. doi: 10.1007/BF02018451

Carston, R., & Uchida, S. (1998). *Relevance theory: Applications and implications*. Amsterdam: John Benjamins.

---

[20] This finding is similar to a trend Gile detected in his research on conference interpreting literature (CIRIN number 33, 2006).

Chesterman, A. (1997). *Memes of translation: The spread of ideas in translation theory*. Amsterdam: John Benjamins.

Endersby, J. W. (1996). Collaborative research in the social sciences: Multiple authorship and publication credit. *Social Science Quaterly, 77*(2), 375-392.

Erwin, K. (2001). *Recherche sur les strategies dans les domains de la traduction et de l'interprétation: Etude scientométrique* (In French) (Unpublished master's thesis). Université Lumière Lyon 2.

Gao, B. (2008). *Conjectures and refutations* (In Chinese) (Unpublished doctoral dissertation). Shanghai International Studies University.

Gile, D. (1995). *Basic concepts and models for interpreter and translator training*. Amsterdam: John Benjamins Pub.

Gile, D. (2000). The history of research into conference interpreting: A scientometric approach. *Target, 12*(2), 297-321.

Gile, D. (2006, number 33). The CIRIN Bulletin [Editorial]. *Http://www.cirinandgile.com/*. Retrieved July 10, 2014, from http://cirinandgile.com/Bulletin%2033rtf.htm

Gile, D. (2006). L'interdisciplinarité en traductologie : Une optique scientométrique. In Interdisciplinarité en traduction (In French) (Ö Kasar, Ed.). In *11e Colloque International sur la Traduction* (pp. 23-37). Istanbul: Isis.

Gile, D. (2010). Chinese BTIs and MTIs: A golden opportunity (W. Zhong, Ed.). In *Proceedings of the 7th National Conference and International Forum on Interpreting* (1st ed., pp. 10-20). Beijing: Foreign Language Teaching and Research Press.

Gile, D. (2013). Institutional, social and policy aspects of research into conference interpreting. In R. Barranco Droege (Author) & O. García Becerra & E. Macarena Prada Macías (Eds.), *Quality in interpreting: Widening the scope* (Vol. 1, pp. 9-31). Granada: Comares.

Gile, D. (2014, number 47). The CIRIN Bulletin [Editorial]. *Http://www.cirinandgile.com/*. Retrieved July 10, 2014, from http://www.cirinandgile.com/Bulletin%2048%20Jul%202014.pdf

Grbić, N., & Pöllabauer, S. (2008). To count or not to count: Scientometrics as a methodological tool for investigating research on translation and interpreting. *Translation and Interpreting Studies, 3*(1), 87-146. doi: 10.1075/tis.3.1-2.04grb

Hall, E. T. (1976). *Beyond culture*. Garden City, NY: Anchor Press.

Hu, G., & Sheng, X. (2000). Interpreting research over the past decade. (In Chinese) *Chinese Science and Technology Translations Journal, 13*(2), 39-44.

Leech, G. N. (1983). *Principles of pragmatics*. London: Longman.

Levelt, W. J., Roelofs, A., & Meyer, A. S. (1999). A theory of lexical access in speech production. *Behavioral and Brain Sciences, 22*(01), 1-38. doi: 10.1017/S0140525X99001776

Li, X. (2007). Trends in China's interpreting research. (In Chinese) *New West, 2*, 206-208.

Lin, Y., Lei, T., Lonergan, J., Chen, J., Xiao, X., Zhuang, H., & Zhang, Y. (1999). *Interpreting for tomorrow: A coursebook of interpreting skills between Chinese and English*. (In Chinese) Shanghai: Shanghai Foreign Education Press.

Liu, M. (2011). Methodology in interpreting studies: A methodological review of evidence-based research. In B. Nicodemus & L. Swabey (Eds.), *Advances in interpreting research: Inquiry in action* (pp. 85-119). Amsterdam: John Benjamins.

Liu, S., & Wang, L. (2007). A survey of interpreting research in China over the past ten years. (In Chinese) *Journal of Guangdong University of Foreign Studies, 18*(1), 37-40.

Lv, G. (2001). Interpreter training needs a scientific teaching plan. (In Chinese) *Chinese Translators Journal, 22*(6), 53-55.

Melin, G., & Persson, O. (1996). Studying research collaboration using co-authorships. *Scientometrics, 36*(3), 363-377. doi: 10.1007/BF02129600

Mu, L., & Wang, B. (2009). Interpreting studies in China: A journal articles-based analytical survey. (In Chinese) *Chinese Translators Journal, 4*, 19-25.

Nasr, M. (2010). *La didactique de la traduction : Une étude scientométrique* (In French) (Unpublished doctoral dissertation). Université Paris 3 Soorbonne Nouvelle.

Pan, J., Sun, Z., & Wang, H. (2009). Professionalization in interpreting: Current development of interpreting in Shanghai and Jiangsu province. (In Chinese) *Journal of PLA University of Foreign Languages, 32*(6), 81-101.

Pöchhacker, F. (1995). "Those who do...": A profile of research(ers) in interpreting. *Target, 7*(1), 47-64. doi: 10.1075/target.7.1.05poc

Pöchhacker, F. (2004). *Introducing interpreting studies*. London: Routledge.

Persson, O., Glänzel, W., & Danell, R. (2004). Inflationary bibliometric values: The role of scientific collaboration and the need for relative indicators in evaluative studies. *Scientometrics, 60*(3), 421-432. doi: 10.1023/B:SCIE.0000034384.35498.7d

Rowbotham, J. (2000). *Enseignement en interprétation et traduction: étude scientométrique d'un échantillon de la littérature* (In French) (Unpublished master's thesis). Université Lyon 2.

Seleskovitch, D. (1978). *Interpreting for international conferences: Problems of language and communication*. Washington: Pen and Booth.

Shanghai International Studies University. (2007). *Job descriptions for R&D at universities* (In Chinese) (pp. 1-170) (China, Ministry of Education).

Snell-Hornby, M. (1995). *Translation studies: An integrated approach* (2nd ed.). Amsterdam: J. Benjamins Pub.

Sperber, D., & Wilson, D. (1986). *Relevance: Communication and cognition*. Cambridge, MA: Harvard University Press.

Stipek, D. J. (1988). *Motivation to learn: From theory to practice*. Englewood Cliffs, NJ: Prentice Hall.

Tang, F. (2010). Empirical research in Chinese interpreting studies. (In Chinese) *Foreign Language World, 137*(2), 39-46.

Tang, S., & Zhou, Y. (1958). The nature of interpreting. (In Chinese) *Western Languages, 2*(3), 321-327.

Wang, B. (2009). Interpreting curriculum design and teaching approaches. (In Chinese) *Journal of Hunan University of Science and Engineering, 30*(3), 208-213.

Wang, E. (2006). Interpretation as a profession in China. (In Chinese) In *Proceedings of the 5th National Conference and International Forum on Interpreting* (pp. 86-97). Shanghai: Shanghai Foreign Language Education Press.

Wen, S. (2006). Simultaneous interpreting for TV programs. (In Chinese) In *Proceedings of the 5th National Conference and International Forum on Interpreting* (pp. 166-177). Shanghai: Shanghai Foreign Language Education Press.

Wittgenstein, L. (1953). *Philosophical investigations*. New York, NY: Macmillan.

Xu, Z. (2014). *The past, present and future of Chinese MA theses in Interpreting Studies: A scientometric survey*. Manuscript submitted for publication, Universitat Rovira i Virgili, Tarragona.

Zeng, J., & Hong, M. (2012). Self-correction in Chinese-English consecutive interpreting. (In Chinese) *Foreign Languages and Their Teaching, 3*(264), 68-71.

Zhan, C., & Suo, R. (2012). Telephone interpreting in China. (In Chinese) *Chinese Translators Journal, 1*, 107-110.

Zhang, W. (1998). *Translating long English sentences into Chinese: An exploratory research* (In Chinese) (Unpublished master's thesis). Beijing Foreign Studies University.

Zhang, W. (2011). A comparative analysis of interpreting studies in China and other countries. *Foreign Languages in China, 8*(5), 94-106.

Zhou, Q. (2007). Simultaneous interpreting in television medium. (In Chinese) *Journal of Fuyang Teachers' College, 3*(117), 58-60.

**Ziyun Xu**

Ziyun Xu is a doctoral researcher at the Intercultural Studies Group of the Universitat Rovira i Virgili. He currently works as Chief Interpreter for the US-China Exchange Council in the United States. In this role he interprets for Chinese and American political leaders, business people and academics, supervises and trains a roster of interpreters and translators, and helps to develop executive training programs in collaboration with Stanford University and the University of California, Berkeley. He also works as a business consultant, facilitating complex negotiations for Chinese start-up companies interested in acquiring clean technologies from US businesses.

*xuziyun@gmail.com*